\documentclass[11pt,preprint]{elsarticle}
\usepackage{txfonts} 
\usepackage{epsfig}
\usepackage{bm}
\usepackage{amssymb}
\usepackage{wrapfig}
\usepackage{ulem}
\usepackage{multirow}
\usepackage{color}

\newcommand{\ksla}{\ooalign{\hfil/\hfil\crcr{k}}}

\newcommand{\cF}{{\cal F}}
\newcommand{\cI}{{\cal I}}

\newcommand{\el}{{\cal L}}
\newcommand{\cM}{{\cal M}}

\newcommand{\psibar}{\mbox{$\overline{\psi}$}}

\newcommand{\vp}{\mbox{$\bm{p}$}}

\newcommand{\vbr}{\mbox{$\bm{r}$}}

\newcommand{\vB}{\mbox{$\bm{B}$}}

\newcommand{\vgamma}{\mbox{$\bm{\gamma}$}}

\newcommand{\bp}{\bf p}

\newcommand{\br}{\bf r}

\newcommand{\nnb}{\mbox{$\nu {\bar \nu}$}}

\begin{document}

\begin{frontmatter}

\title{A Relativistic Quantum Approach to Neutrino and Antineutrino Emission 
via the Direct Urca Process in Strongly Magnetized  Neutron-Star Matter}

\author[nubs,nao,jaea]{Tomoyuki~Maruyama}

\author[WIU,nao]{A.~Baha~Balantekin}

\author[snuv,nao]{Myung-Ki Cheoun}

\author[nao,asTky,BeiH]{Toshitaka~Kajino}

\author[BeiH,nao]{Motohiko~Kusakabe}

\author[ND,nao]{Grant J. Mathews}

\date{\today}

\begin{abstract}
We study  neutrino and  antineutrino emission from the direct Urca process 
in neutron-star matter in the presence of strong magnetic fields.
We calculate the neutrino emissivity of the direct Urca process, 
whereby a neutron converts to a proton, an electron and an antineutrino, or
a proton-electron pair converts to a neutron-neutrino pair. 
We solve exact wave functions for protons and electrons in the states described 
with Landau levels.
We find that the direct Urca process can satisfy the kinematic constraints even 
in density regions where this process could not normally occur in the absence 
of a magnetic field. 
\end{abstract}


\address[nubs]{College of Bioresource Sciences,
Nihon University,
Fujisawa 252-0880, Japan}

\address[nao]{National Astronomical Observatory of Japan, 2-21-1 Osawa, Mitaka, 
Tokyo 181-8588, Japan}

\address[jaea]{Advanced Science Research Center, 
Japan Atomic Energy Agency, Tokai, Naka, Ibaraki 319-1195, Japan}

\address[WIU]{Department of Physics, University of Wisconsin, Madison,
WI 53706, USA}

\address[snuv]{Department of Physics and OMEG Institute, Soongsil University, 
Seoul, 156-743, Korea}

\address[BeiH]{School of Physics, Int. Research Center for Big-Bang 
Cosmology and Element Genesis, Beihang University, Beijing 100083, China}

\address[asTky]{Graduate School of Science, University of Tokyo, Hongo 7-3-1, 
Bunkyo-ku, Tokyo 113-0033, Japan}

\address[ND]{Center of Astrophysics, Department of Physics,
University of Notre Dame, Notre Dame, IN 46556, USA}

\end{frontmatter}

\bigskip

%\maketitle

Neutron stars (NSs) are cooled by neutrino and/or anti-electron neutrino 
emission.
At very low temperatures the neutrino mean-free-path is very long.
In this case neutrinos are easily emitted from the core regions of NSs.
Since neutrino emission rates depend on circumstances inside NSs,
the study of NS cooling through neutrino emission
 gives important information for constraining internal NS structure  
\cite{YP2004}.

There are several kinds of the cooling processes \cite{DNHMNPTEP19}.
Previous works have studied decay processes for the neutrino or the anti-neutrino emission 
such as the direct Urca (DU) process ($n \rightarrow p + e^{-} + {\bar \nu}_e$,
$p + e^{-}  \rightarrow n + \nu_e$),
the modified Urca process (MU) ($n  + N \rightarrow p + e^{-} + N^{\prime} + 
{\bar \nu}_e$,
$p + e^{-} + N \rightarrow n  + N^{\prime} + {\nu}_e$ )
 \cite{URCA},
the neutrino-pair emission process 
($N_1+ N_2 \rightarrow N_1^{\prime} + N_2^{\prime} + \nu + {\bar \nu}$) 
\cite{KPPTY99,OKY14}
and so on.
Furthermore, there are many approaches which  combine the above processes with other 
mechanisms.
For example, one of them is to consider the superfluidity \cite{Lim17} produced 
by neutron and proton pairing correlations. 
This is known to lead to a strong reduction of the neutrino emissivity and affect 
the neutrino-antineutrino emission by the Cooper pair breaking and formation 
mechanism \cite{Wei20}. 

The DU process is one of the most feasible candidates 
to explain the rapid cooling of NSs \cite{Boguta81,LPPH91}.
One usually examines the rapid cooling associated with the DU process to compare 
temperatures and ages of isolated NSs.
Recently another method has been realized by observing NSs in binary 
systems.  
Indeed, by analyzing the x-ray emission in MXB 1659-29, evidence of the direct 
Urca process has been found \cite{BCFJJPR18}

Because of energy-momentum conservation and Fermi statistics  
for NS matter composed of protons, neutrons and electrons,
the DU process occurs in a density region, where the proton density $\rho_p$ is 
larger
than 1/8 of the neutron density $\rho_n$ ($\rho_p \ge \rho_n/8$);
this condition cannot be achieved in low density regions.
However, if the symmetry energy is proportional to the density, and its slope 
is sufficiently large,
the condition for the DU process can be satisfied even at rather low density.
Many other approaches have been proposed to satisfy this condition 
by considering other phases such as pion \cite{MBCDM77} or kaon condensation 
\cite{StC88,FMTT94}  
and matter including hyperon degrees of freedom \cite{Lim17,HypDel92}.
However, the latter case turns out to affect the masses of NSs, which 
is closely related to the hyperon puzzle in NSs \cite{Choi20}. 
%}
  
On the other hand, magnetars, which are associated with a very strong magnetic 
field 
\cite{pac92,mag3}, have properties different from normal NSs. 
The strength of their magnetic field is about $10^{14}-10^{15}$~G in the surface 
region,  
and can reach $10^{17}$~G inside the star. 
Soft gamma repeaters (SGR) and anomalous X-ray pulsars (AXPs) 
correspond to magnetars \cite{Mereghetti08}. 
Magnetars emit energetic photons.   
Furthermore, the surface temperature of magnetars is $T\approx 0.28  - 
0.72~$keV, 
which is larger than those of normal NSs $T \approx 0.01 - 0.15~$keV 
at a similar age \cite{Kaminker09a}. 

There are a number of theoretical works  
that show the influence of the magnetic field on the equation of state (EOS) of NS matter
\cite{FGP89,FGPY92,AS91,RFGPY93,LS91,Ch96,CBP97,BPL00}
and how the structure of NSs is altered by including magnetic 
fields \cite{CPL00,TRYG98} .
We note however, that since our purpose here is to explore effects of the DU 
in NS matter and not the structure of NSs, we do not 
re-examine NS structure in this study. 
For the reader interested in magnetic effects on NS structure a list 
of works can be found in Ref.~\cite{CPL00} (see also the "Lorene" website: {\it 
https://lorene.obspm.fr/data/magnetNS.html}).

A strong magnetic field can supply energy and momentum into the cooling process
and changes the restriction caused by energy-momentum conservation
by introducing transitions between Landau levels \cite{nnbPLB,AxPrd,CV77}.
Indeed, the magnetic field influences the beta-decay process in NSs \cite{LS91}
and axion emission from magnetars \cite{FGSHKS21}.
Leinson \cite{DUwB} studied the effect of magnetic fields perturbatively  
and showed that a magnetic field increases the antineutrino emissivity.
Thus, the associated strong magnetic fields may play a significant role in the 
cooling of magnetars.

In a previous paper \cite{nnbPLB} we have studied \nnb-pair emissions from the 
transition between two Landau levels
in strongly magnetized NS matter.
In that work, when the strength of the magnetic field is about $B \sim 10^{15}$ G,   
emission energies of this process are larger than those of the MU process with 
zero magnetic field, $B=0$.  
All of this suggests that the magnetic field may increase the emission of 
particles 
because the strong magnetic field can supply momentum into the emission 
processes.

Leinson and P\'erez \cite{DUiStMg} calculated 
the neutrino emissivity from the DU process in strong magnetic fields, 
and showed that the emissivity becomes larger as the magnetic field increases.
In that calculation the magnetic field was so strong that all charged particles 
populated the lowest Landau level.
This means that one has not yet estimated realistic magnetic effects in NSs 
with realistic magnetic field strengths.

In the present paper, therefore, we apply our quantum theoretical approach to
the DU process in strong magnetic fields and calculate it through the transition
between Landau levels for electrons and protons. 
As noted above, only this quantum approach can exactly describe the momentum
transfer from the magnetic field.

\bigskip

%\section{Formalism}
%\label{sec-2}

The low energy Lagrangian of the weak interaction between baryons and leptons 
is written as
\begin{eqnarray}
\el_{W} &=& \frac{G_F }{\sqrt{2}} \sum_{l_1, l_2} \psibar_{l_1} \gamma_\mu (1 
- \gamma_5) \psi_{l_2} 
\sum_{\alpha_1 \alpha_2}
\psibar_{\alpha_1} \gamma_\mu (c_V-c_A\gamma_5) \psi_{\alpha_2} ,
\end{eqnarray}
where $\psi_{l_{1.2}}$ is the field of the lepton $l_{1,2}$ which denotes the electron or muon 
or neutrino, and $\psi_\alpha$ is the field of the baryon $\alpha$ which denotes the proton 
or neutron, 
$G_F$ is the Fermi coupling constant, while $c_V$ and $c_A$ are 
the vector and axial-vector coupling constants which include the contribution 
from the Cabbibo angle \cite{rml98}. 

In the single particle model,  
the $l$-neutrino emissivity of the DU process is written as
\begin{eqnarray}
\epsilon_{DU} &=& 2 \sum_{\alpha_n, \alpha_p, \alpha_l, \alpha_\nu} 
\cF_n (E_n, E_p, E_l) 
e_\nu (2 \pi) \delta(E_n - E_p - E_l - e_\nu) 
\nonumber \\ && \qquad 
\times 
\left| G_F \int d^3 \vbr 
\left[ \psibar^{(l)}_{\alpha,_{l}}(\vbr) \gamma_\mu (1 - \gamma_5)  
\psi^{(\nu)}_{\alpha_\nu} (\vbr) \right]  
\left[ \psibar^{(p)}_{n_p}(\vbr) \gamma_\mu (c_V-c_A\gamma_5)  
\psi^{(n)}_{\alpha_n} (\vbr) \right]  \right|^2 ,
\label{DUem}
\end{eqnarray}
with
\begin{equation}
\cF_n (E_n, E_p, E_l) =  n_n (E_n) [ 1 - n_p (E_p) ] [ 1 - n_e (E_l) ]  ,
\end{equation}
where the subscript $l$ indicates the electron or muon,  
while $\nu$ indicates the $\nu_e$ or $\nu_\mu$,
$\psi_a$ is the single particle wave-function, $E_a$ is the energy 
of the particle $a( = n, p, e, \mu, \nu_e, \nu_\mu)$, and $\alpha_a$ denotes 
the quantum numbers  
such as momentum, spin, iso-spin and Landau-number. 
The quantity $n_a (E)$ is the Fermi-distribution function for the particle $a$, 
\begin{equation}
n_a (E) =  \frac{1}{1 + e^{(E-\mu_a)/T} } ,
\end{equation}
where $T$ is the temperature, and $\mu_a$ is the chemical potential of particle 
$a$.

We assume a uniform magnetic field along the $z$-direction,
$\vB = (0,0,B)$, and take the electro-magnetic vector potential $A^{\mu}$ to 
be
$A = (0, 0, x B, 0)$ at the position $\vbr \equiv (x, y, z)$.
The relativistic wave function $\psi$ is obtained
from the following Dirac equation:
\begin{equation}
\left[ \gamma_\mu \cdot (i \partial^\mu - \zeta e A^\mu - U_0 \delta_0^\mu)
- M + U_s
- \frac{e \kappa}{4 M} \sigma_{\mu \nu}
(\partial^\mu A^\nu - \partial^\nu A^\mu ) \right]
\psi_a (x) = 0 ,
\label{DirEq}
\end{equation}
where $\kappa$ is the anomalous magnetic moment (AMM), $e$ is the elementary 
charge
and $\zeta =\pm 1$ or $0$ is the sign of the particle charge.
$U_s$ and $U_0$ are the scalar field and time component of the vector field, 
respectively.
These are determined from relativistic mean-field (RMF) theory \cite{serot97},
although the AMM and the mean-fields are taken to be zero for charged leptons.

The single particle energy for a charged particle ($\zeta=\pm1$) is then written 
as
\begin{equation}
E(n, p_z, s) = E^* + U_0 = \sqrt{ p_z^2 +\left( \sqrt{2 eB n + M^{*2}} 
+ s e  \kappa B /2M \right)^2} + U_0 ,
\label{EsigCh}
\end{equation}
with $M^* = M - U_s$,
where $n$ is the Landau number, 
$p_z$ is the $z$-component of momentum, and $s = \pm 1$ is the spin.
Furthermore, the wave-function overlap at points $\vbr_1$ and $\vbr_2$ for 
charged particles such as the proton and electron 
are written as
\begin{equation}
\psi^{(a)}_{n, s, p_z} (\vbr_1) \psibar^{(a)}_{n, s, p_z} (\vbr_2) = 
\frac{e^{i (p_y y + p_z z)}}{\sqrt{R_y R_z}}
% {\hat F}(\xi_1) \frac{\rho_M}{4 E} {\hat F}(\xi_2)
 {\hat F} \left(x_1 - p_y / eB \right) \frac{\rho^{(a)}_M}{4 E} 
 {\hat F} \left(x_2 - p_y / eB \right) 
\end{equation}
with 
%$\xi_{1,2} = x_{1,2} - \zeta p_y /eB$, and  
%
\begin{eqnarray}
\rho_M (n, s, P_z) &=&  
\left[ E^* \gamma^0 - \zeta \sqrt{2eBn}\gamma^2 - p_z \gamma^3 
+ M^* + \frac{e \kappa B}{2 M} \Sigma_z \right] 
%%%
\nonumber \\&&  \qquad \times
\left[ 1 + \frac{s}{\sqrt{ 2eB n  + {M^*}^2}} \left(
\frac{e \kappa B}{2M}
 + p_z \gamma_5 \gamma^0 -  E^* \gamma_5 \gamma^3 \right) \right] ,
%%%%
\nonumber \\
{\hat F} &=& {\rm diag} \left( f_{n}, f_{n-1}, 
f_{n}, f_{n-1} \right)
= f_{n}  \frac{1 + \Sigma_z }{2}
+  f_{n-1}  \frac{1 - \Sigma_z }{2}  \quad ({\rm proton}) ,
\\
{\hat F} &=& {\rm diag} \left( f_{n-1}, f_{n}, 
f_{n-1}, f_{n} \right)
= f_{n-1}  \frac{1 + \Sigma_z }{2}
+  f_{n}  \frac{1 - \Sigma_z }{2}   \quad ({\rm electron}) ,
\end{eqnarray}
where
 $\Sigma_z = {\rm diag}(1,-1,1,-1)$, and  $R_a$ ($a=x, y, x$) is the size of
the system along the $a$-direction.

For neutral particles ($\zeta = 0$) such as the neutron, 
the wave function is written as an eigenstate of  momentum $\vp \equiv (\vp_T, 
p_z)$.
So, the single particle energy is given by
\begin{equation}
E(\vp, s) = E^* + U_0 = \sqrt{ p_z^2 + \left( \sqrt{ \vp_T^2 + M^{*2}}
+ s e  \kappa B /2M \right)^2 } + U_0 ,
\label{EsigNt}
\end{equation}
and the  wave-function overlap becomes
%
%========================
 \begin{eqnarray}
\psi_{\bp, s} (\vbr_1) \psibar_{\bp, s} (\vbr_2) &=&
\frac{e^{i \bp \cdot (\br_1 - \br_2)}}{\sqrt{R_x R_y R_z}}
\left[ E^* \gamma_0 - \vp \cdot  \vgamma  
+ M^* + \frac{e \kappa B}{2 M} \Sigma_z \right] 
%%%
 \nonumber \\&&  \qquad \times
\left\{ 1 + \frac{s}{\sqrt{ {\bf p}_T^2  + {M^*}^2}} \left[
\frac{e \kappa B}{2M} + \gamma_5 \left(
 p_z \gamma^0 -  E^* \gamma^3 \right) \right] \right\} .
%%%%
\end{eqnarray}

Substituting the above wave-functions into Eq.~(\ref{DUem}), we obtain
\begin{eqnarray}
\epsilon_{DU} &=& 
\frac{2 (2 \pi)^3  G_F^2 (eB)^2}{ R_x^2 R_y^4 R_z^4 } \sum_{n_p, n_e} \sum_{s_n, 
s_p s_e}  
 \left\{ \prod_{a = n, \nu} \left[ R_x R_y R_z \int \frac{d^3 p_a}{(2 \pi)^3} 
\right] \right\} 
\left\{ \prod_{b=p, e} \left[ R_z \int \frac{d p_{bz}}{2 \pi} 
  R_y \int^{eBR_x/2}_{-eBR_x/2} \frac{d p_{by}}{2 \pi} \right] \right\} 
\nonumber \\ && \quad \times 
e_\nu \cF_n (E_n, E_p, E_{l}) 
\delta (E_n - E_p - E_{l} - e_\nu) \delta (p_{ny} - p_{py} - p_{l y} - p_{\nu 
y} )
   \delta (p_{nz} - p_{pz} - p_{l z} - p_{\nu z} )
\nonumber \\ && \quad
\times \frac{1}{16^3 E_n E_p E_{l} e_\nu}  
\sum_{i_1,  j_1,  i_2,  j_2 (=\pm 1)} \cM(j_1, i_1)  \cM^* (j_2, i_2) L_{\mu 
\nu} N^{\mu \nu}
%--------------------------------------
\nonumber \\ &=& 
 \frac{G_F^2(eB)^3}{4 (2 \pi)^5} \sum_{n_p, n_l} \sum_{s_n, s_p, s_l}
 \int \frac{dp_{nT} p_{nT} d p_{pz} d p_{l z} d e_\nu e_\nu^3}{E^*_n E^*_p E_{l} 
e_\nu} 
\frac{d \Omega_\nu}{4 \pi} \cF_n ( E_n, E_p, E_{l} )
\nonumber \\ && \qquad
\times  \delta(E_n - E_p - E_{l} - e_\nu)   
\sum_{i_1, j_1, i_2,  j_2 } \cM(j_1, i_1)  \cM^* (j_2, i_2) L_{\mu \nu} (i_1, 
i_2) N^{\mu \nu} (j_1, j_2) ,
\label{DUMag1}
\end{eqnarray}
%
%====================================
with 
\begin{eqnarray}
\cM  (j_p, j_{l})  &=&
\int d x f_{n_e+ (j_{l}-1)/2} \left(x + \frac{p_{n T}}{\sqrt{2eB}} \right)  
f_{n_p+ (j_p-1)/2} \left(x - \frac{p_{n T}}{\sqrt{2eB}} \right),
\label{OvFn}
\\
L_{\mu \nu} (j_1, j_2) &=& \frac{1}{16} {\rm Tr} \left\{ 
\rho_M^{(e)}(n_f,s_f, k_{e})  (1 + j_1 \Sigma_z)
 \gamma_\mu (1 - \gamma_5)
\ksla_{\nu} \gamma_\nu (1 - \gamma_5) (1 + j_2 \Sigma_z) \right\} ,
%\nonumber 
\\
%%%%%%%%%%
N_{\mu \nu} (j_1, j_2) &=&  \frac{1}{16} {\rm Tr} \left\{
 (1 + j_2 \Sigma_z) 
\rho_M^{(p)}(n_i,s_i,P_{iz})  (1 + j_1 \Sigma_z)
\gamma_\mu (c_V - c_A \gamma_5) 
\right. \nonumber \\ && \qquad\quad  \times \left.
\rho_M^{(n)} \gamma_\nu (c_V - c_A \gamma_5) \right\} ,
\end{eqnarray}
where $R_x = R_y = R_z$ is assumed, and $f_n$ is the $n$-th harmonic oscillator 
wave function.
In addition, the neutron momentum is taken to be 
$p_n =(E_{n}, 0, p_{nT}, p_{nz})$, where without loss of generality 
the neutron transverse momentum is assumed to be directed along the $y$-axis.

In cool NSs the temperature is less than 1 keV, and the emitted neutrino energy 
is 
of the order of the temperature $e_\nu \sim T$, 
so that we can take the lowest order term of the temperature 
by using the following approximation
\begin{equation}
e_\nu^3 \cF_n (E_n, E_p, E_{l})   
\approx \cI_{DU} \delta(E_n - \mu_n) \delta(E_p - \mu_p) \delta(e_e - \mu_e) ,
\end{equation}
with 
\begin{equation}
\cI_{DU} = \int d E_1 d E_2 d E_3 (E_1 - E_2 - E_3)^3 \cF_n (E_1, E_2, E_3) 
\approx \frac{457}{5040} \pi^6 T^6 .
\end{equation}

Therefore, the neutrino emissivity in the low temperature limit can be written 
as
\begin{eqnarray}
\epsilon_{DU} &= &
\frac{457 \pi }{10080} G_F^2 T^6 \sum_{n_l , n_p} 
\frac{ p_{nT} }{ p_{pz} p_{lz} \sqrt{p_{nT}^2 + {M_n^*}^2}}  
 \nonumber \\ &&  \qquad \qquad\qquad \times 
\frac{1}{2^6}
 \sum_{s} \sum_{i, j} \int \frac{d \Omega_\nu}{4 \pi} \cM(j_1, i_1)  \cM^* (j_2, 
i_2) 
\frac{ L_{\mu \nu} N^{\mu \nu} }{e_\nu}.
\label{emiB}
\end{eqnarray}

In addition, we assume chemical equilibrium for the system.
Hence, the chemical potentials satisfy the following relations:
\begin{equation}
\mu_n = \mu_n^* + U_0 (n) = \mu_p + \mu_l = \mu_p^* + \mu_l + U_0(p) . 
\end{equation}
At the low temperature limit, the chemical potentials are taken to be the Fermi 
energies at zero temperature,
and the effective chemical potentials for baryons $\mu^*$  in the presence of 
a magnetic field are written as
\begin{eqnarray}
\mu_p^* (n_p, k_{pz}, s) &=& \sqrt{ \left( \sqrt{ 2eB n_p + {M_p^*}^2 } + s e 
\kappa_p / 2 M_p \right)^2 + p_{pz}^2 } ,
\\
\mu_n^* (k_{nT}, k_{nz}, s) &=& \sqrt{ \left( \sqrt{ k_{nT}^2 + {M_n^*}^2 } + 
s e \kappa_n / 2 M_n \right)^2 + k_{nz}^2 }  ,
\end{eqnarray}
where $s=\pm 1$, $n_p$ is the Landau number of the proton, and $k_{pz}$ is the  
$z$-component of the proton Fermi momentum 
which depends on $n_p$.
In addition, $k_{nT}$ and $k_{nz}$ are the transverse component and 
$z$-component of the neutron  Fermi momentum,
whose values are not the same.
That  is  the momentum-distribution breaks the spherical symmetry through
the AMM. 

For comparison we also show the neutrino emissivity from the DU process 
when $B=0$ as
\begin{eqnarray}
\epsilon_{DU}    &=&
\frac{457 \pi}{10080} G_F^2 T^6
\left\{ \frac{(c_V - c_A)^2}{2} \mu^*_p  
\left( {M_n^*}^2 - {M_p^*}^2 + m_l^2 - 2 \mu^*_p \Delta U_0  - \Delta U_0^2  
\right)
\right. \nonumber \\ && \left. \quad
+ \frac{(c_V + c_A)^2}{2} \mu^*_n 
\left( {M_n^*}^2 - {M_p^*}^2 - m_l^2 - 2 \mu^*_n \Delta U_0 + \Delta U_0^2  
\right)
-  (c_V^2 - c_A^2) \mu_l M^*_p M^*_n  \right\} ,
\end{eqnarray}
where $\Delta U_0 = U_0(p) - U_0(n)$;
this expression is equivalent to that of Ref.~\cite{DUinR1}.
%
%\textcolor{red}{
This emissivity is non-zero $\epsilon_{DU} > 0$ when $p_F(n) \le p_F(p) + 
p_F(l)$  with $p_F(a)$,  the Fermi momentum of the particle $a$.
This momentum matching condition determines the density region for the DU process 
when $B=0$.
In contrast, for $B \neq 0$, the magnetic field carries momentum and breaks the 
momentum matching condition, and the DU process can occur
in any density region.  
%}

In this work we discuss the relation between the DU emissivity and the 
density dependence of the symmetry energy as well as the effects of the magnetic field.  
For the calculation of nuclear matter, we use the following Lagrangian density  
in the relativistic mean-field approach:
\begin{eqnarray}
 \el_{RMF} 
&=&
\psibar \left( i \gamma \cdot \partial - M + g_s \sigma \right) \psi
% \nonumber \\ &&
+ \tilde{U} [\sigma] + \frac{g_V^2}{2 m_V^2} \left( \psibar \gamma \psi 
\right)^2
% \nonumber \\ &&
- \frac{C_s^{IV}}{2 M^2} \left( \psibar \tau \psi \right)^2  
- \frac{C_v^{IV}}{2 M^2} \left( \psibar \gamma \tau \psi \right)^2  ,
\end{eqnarray}
where $\tilde{U} [\sigma]$ is the self-energy term of the sigma meson which 
includes
the non-linear effect for the scalar mean-field,  and
the other terms are written as the zero-range interaction between two nucleons.
In addition,  $C_{s}^{IV}$ and $C_{v}^{IV}$  are the coupling constants for the 
iso-vector interaction in the Lorentz scalar and vector channels, respectively 
\cite{RMF-SEn}. 
Then, the scalar-field $U_s$ and the time component of the vector-field are given 
by
\begin{eqnarray}
U_s &=& g_s \sigma + \frac{C_s^{IV}}{2 M^2} (\rho_s (p) - \rho_s (n) ) \tau_z 
~~,
\\
U_0 &=& \frac{g_V^2}{m_V^2} ( \rho_p + \rho_n )  
+ \frac{C_v^{IV}}{2 M^2} (\rho_p - \rho_n ) \tau_z .
\end{eqnarray}

In the above formalism we omit the magnetic field energy density 
$B^2/(8 \pi)$ in the EOS. This term  contributes to the pressure and affects the  composition of NS matter
significantly  when $B \gtrsim 10^{18}$ G \cite{BPL00}.
In the present calculation we adopt a smaller  strength of the magnetic field, and this term does not
contribute to the neutrino emissivity.

\begin{table}[h]
% \vspace{0.5cm} 
%\begin{wraptable}{r}{8.2cm}
\begin{center}
\begin{tabular}{|c|c|c|c|}
\hline 
Parameter-Set & ~$C_s^{IV}$~  & $C_v^{IV}$ & $L$~(MeV) \\
\hline
SF1 & 0 & 20.42 & 92.7  \\
\hline
SF2 & 23.61 & 0 & 84.1 \\
\hline
SF3 & 33.02~ & -8.154 & 81.0 \\
\hline
\end{tabular}
\caption{\small
The parameter-sets for the iso-vector parts  of the mean-fields.
All parameter-sets give a symmetry energy of 32~MeV at normal nuclear density.
The fourth column shows the slope parameter of the symmetry energy. 
}
\end{center}
\label{SFpara}
%\end{wraptable}
\end{table}
%\vspace{0.5cm}

\begin{wrapfigure}{r}{7.8cm}
\vspace*{-1.em}
%\begin{figure}[htb]    
\begin{center}
\includegraphics[scale=0.35,angle=270]{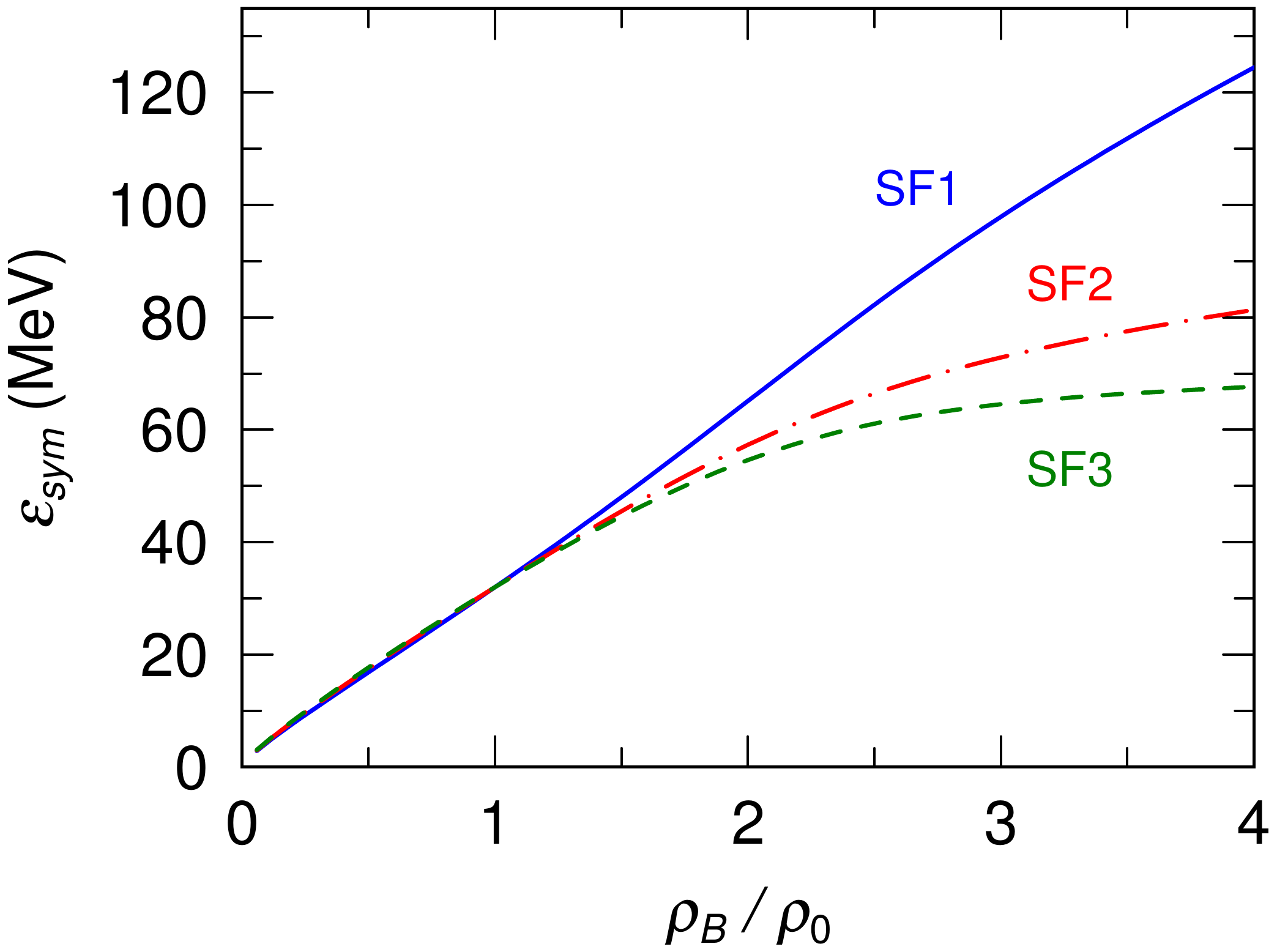}
\caption{
The density dependence of the symmetry energy.
The solid, dot-dashed and dashed lines represent the results
with the parameter sets SF1, SF2 and SF3, respectively.}
 \label{SymEn}
\end{center}
%\end{figure}
\end{wrapfigure}

We use the parameter-set PM1 for symmetric matter as given in Ref.~\cite{K-con}.
In addition, we give the three parameter-sets, SF1, SF2 and SF3, 
for the other parameters of the iso-vector parts, 
$C_s^{IV}$ and $C_v^{IV}$, whose detailed values are shown in 
Table~1 . %\ref{SFpara}.
The symmetry energy is fixed to be $e_{sym} = 32$~MeV at normal nuclear density
though the parameter-sets give different density dependences of the symmetry 
energy.  
The SF1 includes only the Lorentz vector channel ($C_s^{IV}=0$), 
the SF2 includes only the iso-vector Lorentz scalar channel  ($C_v^{IV}=0$), 
and the SF3 includes the negative value of the Lorentz vector channel 
($C_v^{IV} < 0 < C_s^{IV}$).  
In addition, we list the values of the slope parameter $L$ for these 
parameter-sets.

%AAAA
In Fig.~\ref{SymEn} we show the density-dependence of the symmetry energy, 
We see that the three parameter-sets give quite different density dependences 
of 
the symmetry energy though all of them satisfy 32 MeV for the symmetry energy 
at normal nuclear density.
This means that the parameter-sets, SF1, SF2 and SF3,  represent
different EOSs.

In actual calculations we substitute the mean-fields and the chemical potentials 
at $B=0$ with these parameter-sets into Eq.~(\ref{emiB}) and calculate the 
neutrino emissivity.
As mentioned above, surface temperatures of magnetars are 
$T\approx 0.28  - 0.72~$keV \cite{Kaminker09a}.
Thus, we choose $T = 0.50$~keV for the present calculations.

\begin{figure}[htb]    
%\vspace*{2em}
\begin{center}
\includegraphics[scale=0.6]{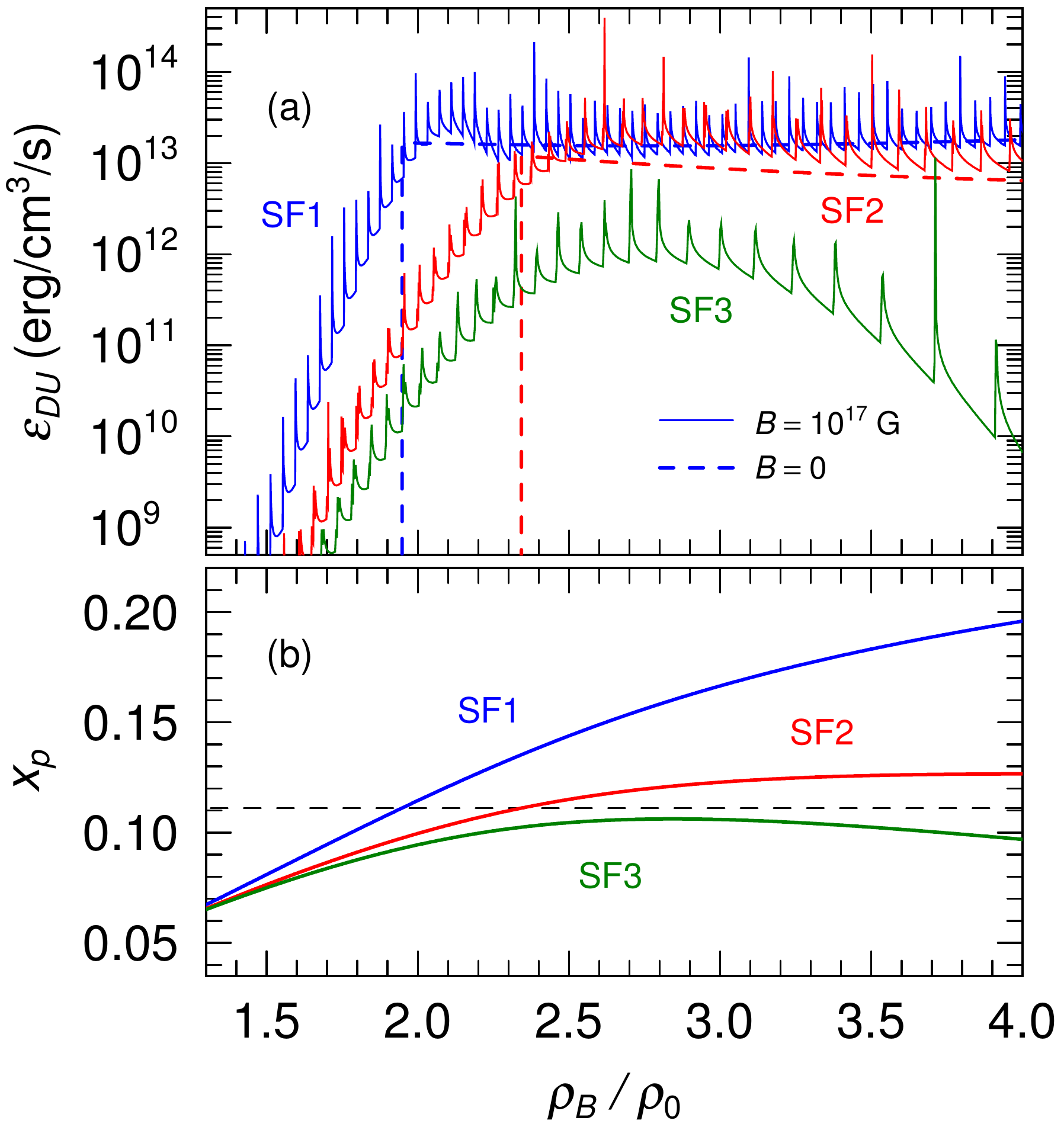}
 \caption{
(a) The density dependence of the neutrino emissivity in the DU process at 
$T=0.50$~keV 
 and (b) that of the proton fraction of NS matter at $B=0$.
The vertical lines represent the results with the SF1, SF2 and SF3, respectively.
 In the upper panel (a) the solid and dashed lines indicate the results when 
$B = 10^{17}$G 
 and $B = 0$, respectively.
 Note that the dashed line for the SF3 is not shown because $\epsilon_{DU} =$0.}
 \label{DUPM1}
\end{center}
\end{figure}

First, we restrict the matter to consist of only protons, neutrons and electrons.
In Fig.~\ref{DUPM1} we show the density dependence of the neutrino emissivity 
in the DU process (a)
and the density dependence of the proton fraction in NS matter, 
$x_p$ (b).
In Fig.~\ref{DUPM1}(a), the dashed line indicates the results for $B=0$, which  appear 
in the density region of $x_p \ge 1/9$;  $x_p=1/9$ is denoted by the horizontal 
dashed line  
in the lower panel (b). 
For future reference, we define $\rho_{DU}$ as the critical density 
at which the proton fraction $x_p = 1/9$. 

The calculations result in large fluctuations, which also appears the in 
\nnb-pair \cite{nnbPLB} and axion emission \cite{AxPrd}.
These fluctuations reflect the density of states in the $xy$-plane given by the 
Landau levels \cite{CV77, QH}.

%\bigskip

\begin{wrapfigure}{r}{8cm}
\vspace*{-0.7em}
\begin{center}
\includegraphics[angle=270, scale=0.48]{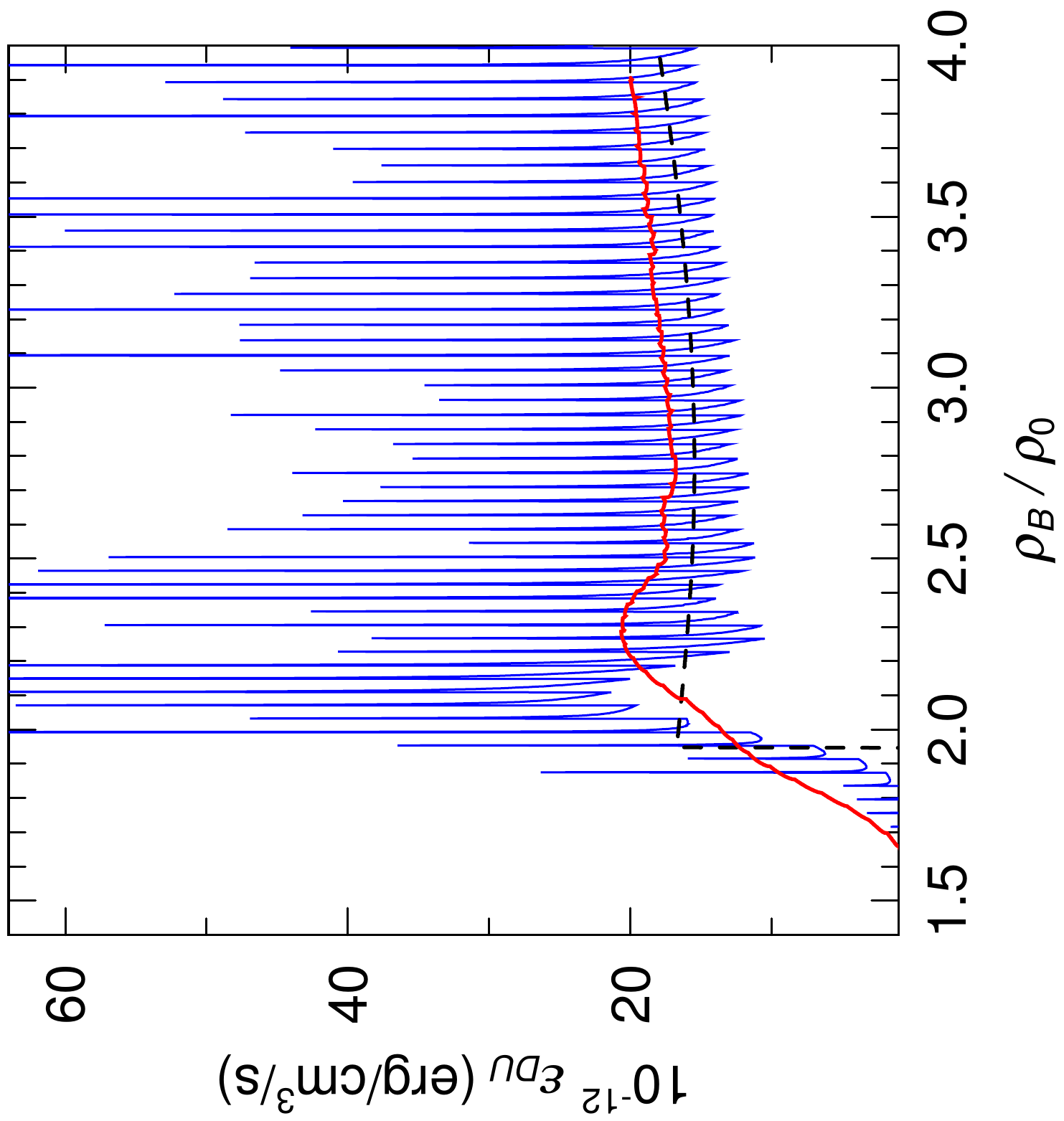}
 \caption{
 The density dependence of neutrino emissivity in the DU process at $T=500$~eV 
and $B=10^{17}$~G (a)
for PM1-SF1.
The thin solid blue and dashed lines indicate the same quantities as those in 
Fig.~\ref{DUPM1}a 
while the scale is linear.
The  thick solid red line represents averaging over the density dependence of 
the emissivities (thin solid blue line) using a width of  $0.01$ fm$^{-3}$.
 }
 \label{DUPM1Ls}
\end{center}
\end{wrapfigure}

When $B=0$, the results of the neutrino emissivity for SF1 and SF2 suddenly appear 
at the density $\rho_{DU}$.  
They increase very rapidly  and become almost flat as the density increases.
When $B=10^{17}$G, the density dependence of the neutrino emissivity has 
comb-like-shapes  
which are due to the fact that the neutrinos are emitted through transitions 
between Landau  levels.
The emissivity appears at a density of $\rho < \rho_{DU}$ and becomes larger 
gradually
with increasing density.
In the density region $\rho \ge \rho_{DU}$, they undergo large oscillations  
though their local minima almost agree with the $B=0$ results.

For SF3, on the other hand, the proton fraction does not exceed $1/9$,
and the DU does not occur when $B=0$.
When $B=10^{17}$G, however, neutrinos are emitted, and
thus, we can confirm that the magnetic-field increases the neutrino emission.
In addition, the peak position of the neutrino emissivity agrees with that of 
the proton fraction,
namely the effect of the magnetic field is related to the proton fraction and 
can appear when the proton fraction is close to $1/9$.

In Fig.~\ref{DUPM1Ls} we plot the results for SF1 on a linear scale.
Here one can see that the spike of the neutrino emissivity is very narrow, 
and that its height is much larger than that with zero-magnetic field.
To estimate the strength of the neutrino emission, 
we smooth the emissivity according to the following equation:
\begin{equation}
{\bar \epsilon}_{DU} (\rho_B) = \frac{1}{\Delta \rho} 
\int_{\rho_B - \Delta \rho/2}^{\rho_B + \Delta \rho/2} d \rho \epsilon_{DU} 
(\rho).
\end{equation}
We show this result for SF1 when $\Delta \rho = 0.01$~fm$^{-3}$ with a solid 
line.
 The smoothed value for the emissivity is still larger than that when $B=0$, 
 but the difference is not very drastic.

\begin{wrapfigure}{r}{8.2cm}
%\begin{figure}
%\vspace*{-0.5em}
\begin{center}
\includegraphics[scale=0.51]{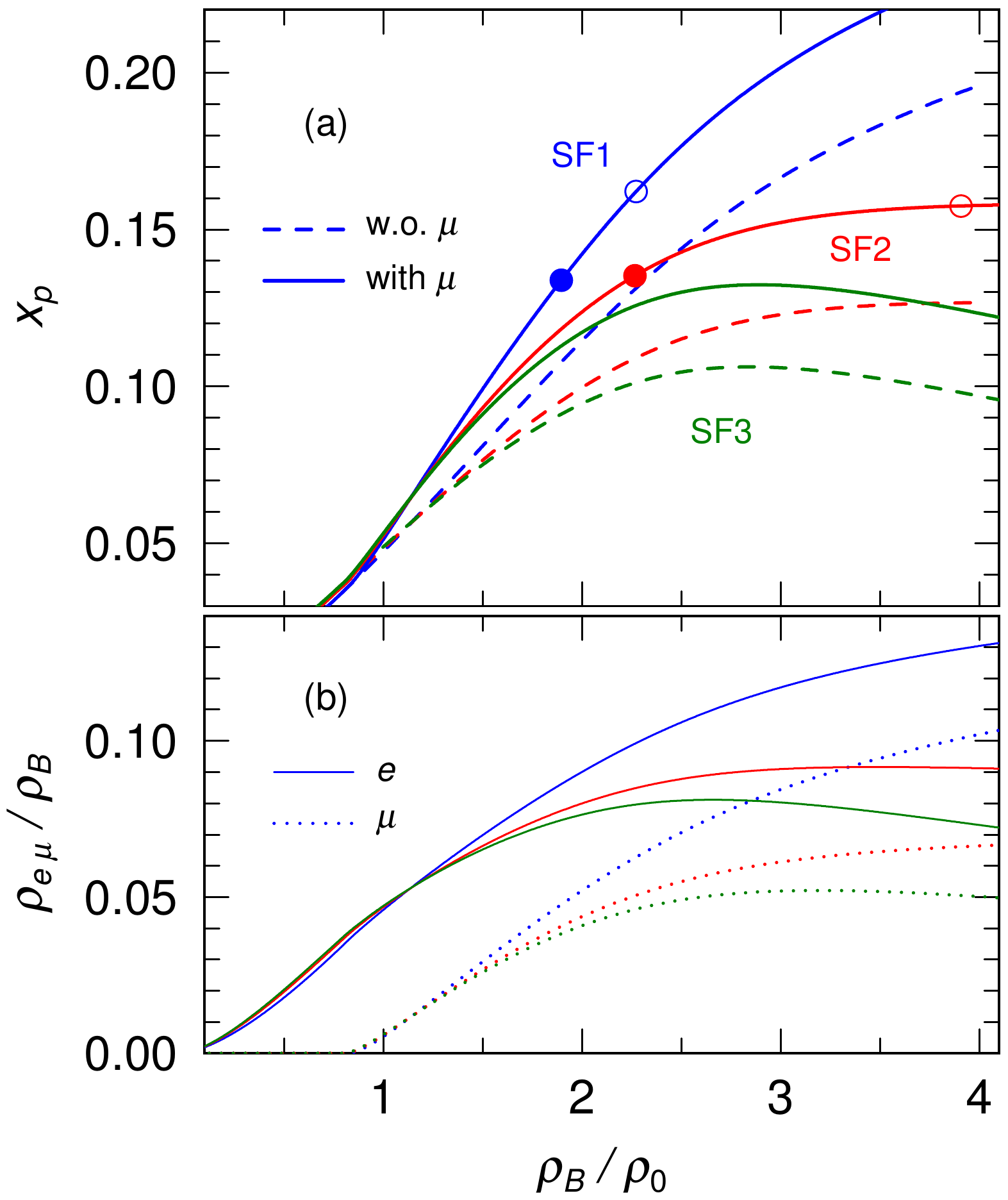}
 \caption{
Proton fractions (a)  and lepton fractions (b) in NS matter with  muons.
In the upper panel the solid and dashed lines represent the proton fractions 
in NS matter with and without muons, respectively.
In the lower panel the thin solid and dotted lines represent the electron and 
muon fractions.
Lines represent the results from  the SF1, SF2 and SF3 EoS as 
labeled.
}
 \label{PrFrMu}
\end{center}
\end{wrapfigure}
%\end{figure}
%

Here, we add the following comment.
In the study of pion \cite{P2Pi-1, P2Pi-2}, axion \cite{AxPrd} and \nnb-pair 
production \cite{nnbPLB}
we showed a very large difference in the results between cases including 
the AMM of the protons and those not including it.
In this calculation there is no significant difference between the cases with 
and without the AMM.
Without a magnetic field the above particle production is forbidden by 
energy-momentum conservation.
Thus,  only the argument of the overlap function of Eq. (\ref{OvFn}) in the tail 
region contributes 
to the transition strength, and a small difference in input values makes a very 
large difference 
in the calculation results.
In the DU process the transitions are not completely forbidden by the kinematics, 
and 
the value of the argument $p_{nT}/\sqrt{2eB}$
is not located in the tail region.

%\textcolor{red}{
As a next step we study the DU process in NS matter including muons.
In Fig.~\ref{PrFrMu} we show the proton fraction in such matter in the upper 
panel
and the electron and muon densities per baryon in the lower panel.
The solid and open circles show the critical densities of $\rho_{DU}$ 
for the electron and muon neutrino emission, respectively.
In matter containing muons the proton fraction becomes larger, and the critical density 
$\rho_{DU}$ becomes a little lower than in  matter without muons. 
Here, we note that  the proton fraction in the SF3  EOS is also larger and  
exceeds $1/9$, but the DU does not appear.
%}

\begin{figure}
\vspace*{-0.5em}
\begin{center}
\includegraphics[angle=270,scale=0.6]{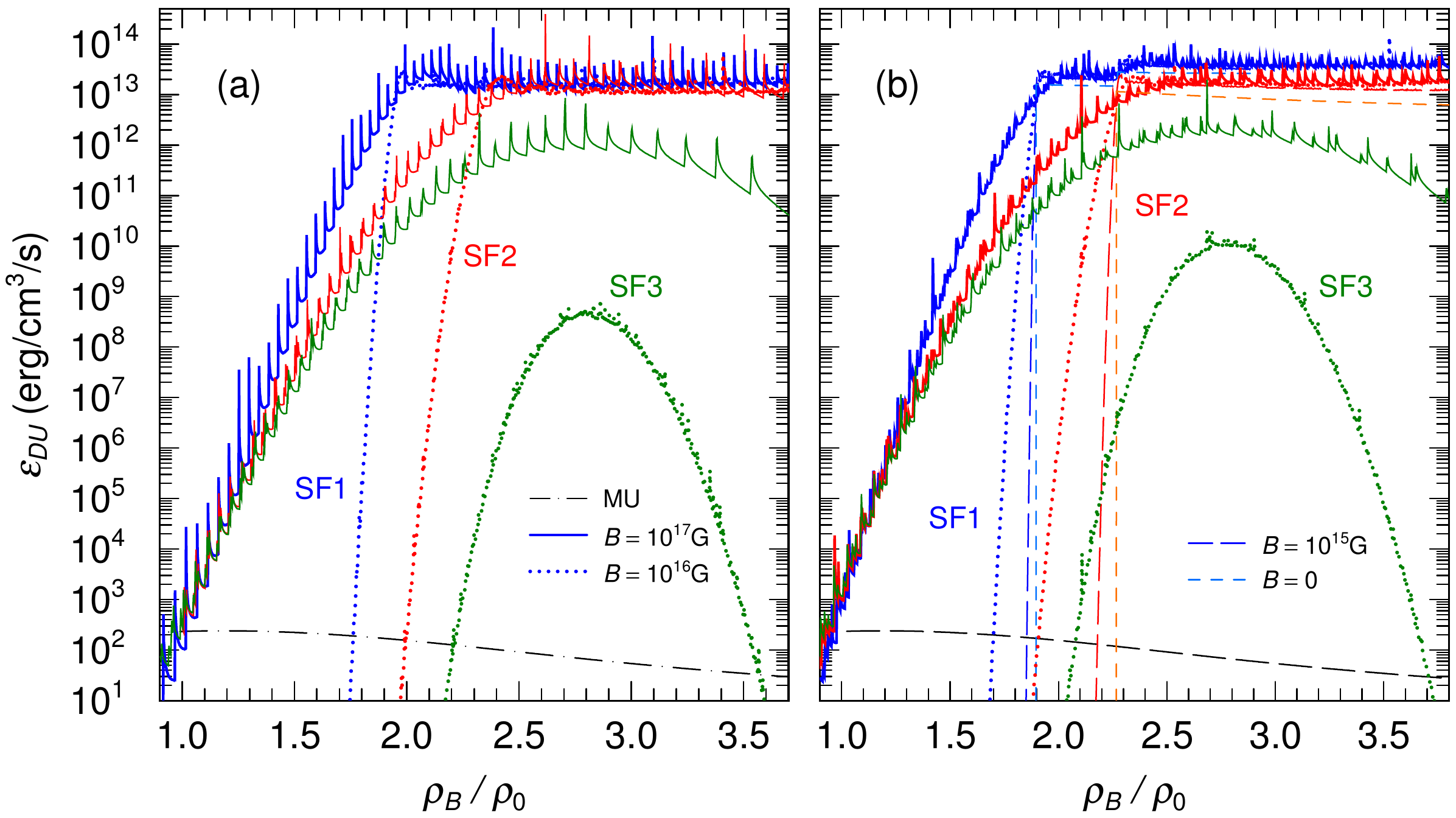}
 \caption{
 Density dependence of the neutrino emissivity in the DU process at $T=500$~eV for  the PM1-SF1 EoS and for 
 matter containing only protons, neutrons and electrons (a) and  matter  also containing  muons  (b).
The solid, dotted and long-dashed lines represent the results 
 for magnetic field strengths, $B=10^{17}$, $10^{16}$ and $10^{15}$~G, respectively, and
the dashed lines indicate the results without a magnetic field.
The long-dashed line shows the neutrino emissivity from the MU process.
 }
 \label{DUBdp}
\end{center}
\end{figure}

In order to examine the dependence on the magnetic field strength, in 
Fig.~\ref{DUBdp} we show the results at
$B=10^{17}$, $B=10^{16}$ and  $B=10^{15}$~G for
 NS matter without muons (a) and with muons (b). 
The results for SF3 when $B=10^{15}$~G  are too small to appear in this figure.
For comparison, we also show the neutrino luminosities
from  the MU process \cite{MURCA} with the long-dashed lines.
When $B \lesssim 10^{15}$~G, the emissivities do not appear for $\rho_B < \rho_{DU}$,
and the magnetic field does not play an important role for the DU process.

The emissivities from the NS matter with muons are larger than those without muons.
Particularly, the emissivity  for SF3  at $B=10^{16}$~G is much larger in the matter with muons 
than that without muons
though the contribution from the muon process is negligibly small.

As the strength of the magnetic field decreases, the amplitudes of the 
oscillations 
in the density dependence become smaller for both magnetic field strengths in NS matter.
In the density region, $\rho_B < \rho_{DU}$,  
the neutrino emissivities are largely suppressed, and the DU process dominates 
in a narrower density region, 
as shown in the results by SF3, 
where the emissivities in the DU process are still much larger than that of the 
MU process.

Here, we observe the following facts:
As the magnetic field strength decreases,  the emission rate becomes larger for 
the \nnb-pairs~ \cite{nnbPLB}  
despite a smaller momentum transfer from the magnetic field.
The \nnb-pairs are produced via the transition of the proton and electron
between different  Landau levels;
namely the initial and final particles are the same.
As the magnetic field strength decreases,  the energy interval between
the initial and final states also becomes smaller.
Hence, the emission strength becomes larger.
In the DU process, however, the initial particle is a neutron which does not 
stay in the Landau level, 
and the energy interval is continuous, so that this effect does not appear. 

In summary, we have used a relativistic quantum approach 
to study neutrino and anti-neutrino  emission in the DU process from NS matter 
with  strong magnetic fields, $B = 10^{15} - 10^{17}$~G.
We use  three parameter-sets for the symmetric nuclear force 
to illustrate the relation between the proton fraction and the neutrino 
emissivity.
If the proton fraction satisfies the DU condition $x_p \ge 1/9$, 
the neutrino emissivities are not much different from the case of $B=0$, 
and the magnetic field does not significantly amplify the emission, 
though it causes very large fluctuations in the density dependence of the 
neutrino emissivity.
In the usual forbidden region $x_p < 1/9$, however, the magnetic field 
contributes to the emission and
changes the kinematical condition.
The effect is larger when the magnetic field strength is larger and the proton 
fraction is closer to $1/9$. 

If the magnetic field is rather weak $B \lesssim 10^{15}$~G,  the DU process 
does not appear when $x_p < 1/9$,
so that it does not occur in the surface region of NSs.
However,   neutrino emission from the Landau quantization may occur 
in the inner core of magnetars which are thought to have such strong magnetic 
fields.
Those neutrinos are able to escape from the matter 
because they have very low energy, and their mean-free-paths are very long.

Magneto-hydrodynamic proto-NS simulations \cite{BrSp04,TKS09,KM10} have 
demonstrated  
that the magnetic field inside a NS can obtain a  toroidal 
configuration.
It has also been demonstrated~\cite{TKS09} that the field strength
of toroidal magnetic fields can be $\sim$100 times stronger
than that of a poloidal magnetic field.
Indeed, by analyzing the hard X-ray data and precession of pulsars.
Makishima et al. \cite{Makishima14} suggested evidence for the 
existence 
of  a toroidal magnetic field whose
strength is about one hundred times that of  poloidal magnetic fields for cold 
NSs.

In the present calculation the neutrino emissivities vanish  when $x_p \ll 1/9$ because 
we use the low temperature approximation 
whereby the Fermi distribution functions are approximated by a  step function. 
As mentioned above,  we exactly calculated the \nnb-pair  production in  strong 
magnetic fields \cite{nnbPLB}
and showed that the neutrino emissivity is larger than that of  the MU process 
in the case of  zero magnetic field, even though the  production rates described 
here are zero 
in the low temperature approximation. 
This implies that  there may be much larger neutrino emissivities 
in the DU process when $x_p \ll 1/9$; it may be possible to explicitly show that 
with an exact calculation. 
We defer exploring this point to a future work

Our final remark is about the EOS as there are many different versions from many 
nuclear models. 
A salient feature is the density-dependence of the symmetry energy, 
which is related to the proton fraction in NS matter.
One needs to explore using different EOSs \cite{Ch13,RXCRS11} 
which we also defer to a future work.

\medskip
This work was supported in part by the Grants-in-Aid for the Scientific Research 
from the Ministry of Education, Science and Culture of Japan 
(JP20K03958, JP19K03833, JP17K05459, JP16K05360, JP15H03665). 
ABB is supported in part by the U.S. National Science Foundation
Grants No. PHY-1806368, PHY-2020275, PHY-2108339. 
MKC is supported by the National Research Foundation of Korea 
(Grant Nos. NRF-2021R1A6A1A03043957 and 2020K1A3A7A09080134, NRF-2020R1A2C3006177). 
Work of GJM supported in part by the U.S. Department
of Energy under Nuclear Theory Grant DE-FG02-95-ER40934.

\end{document}